\documentclass[epj]{svjour}
\usepackage{graphics}
\usepackage{subfigure}
\begin{document}
\title{Centrality dependence of $v_2$ in Au + Au at $\sqrt{s_{NN}}$ = 200 GeV}
\subtitle{
  unidentified charged hadron $v_2$ with respect to the first harmonic ZDC-SMD event plane
}
\author{Hiroshi Masui\inst{1} for the PHENIX Collaboration 
}                     
%
%
\institute{Lawrence Berkeley National Laboratory, Nuclear Science Division, 1 Cyclotron Road, Berkeley, CA 94720}
%
\date{Received: date / Revised version: date}
%
\abstract{
    One of the most striking results is the large elliptic flow ($v_2$) at RHIC.
  Detailed mass and transverse momentum dependence of elliptic flow are well 
  described by ideal hydrodynamic calculations for $p_{\mathrm{T}} < $ 1 GeV/c, 
  and by parton coalescence/recombination picture for $p_{\mathrm{T}} = 2 - 6$ GeV/c.
  The systematic error on $v_2$ is dominated by so-called "non-flow effects", 
  which is the correlation not originated from reaction plane. 
  It is crucial to understand and reduce the systematic error from non-flow effects 
  in order to understand the underlying collision dynamics.
    In this paper, we present the centrality dependence of $v_2$ with respect to the 
  first harmonic event plane at ZDC-SMD ($v_2$\{ZDC-SMD\}) in Au + Au collisions 
  at $\sqrt{s_{NN}}$ = 200 GeV.  Large rapidity gap ($|\Delta\eta| > 6$) between 
  midrapidity and the ZDC could enable us to minimize possible non-flow contributions.  
  We compare the results of $v_2$\{ZDC-SMD\} with $v_2$\{BBC\}, which is measured by 
  event plane determined at $|\eta| = 3.1 - 3.9$. 
  Possible non-flow contributions in those results will be discussed.
\PACS{
      {25.75.-.q}{Relativistic heavy-ion collisions}   \and
      {25.75.Ld}{Collective flow}
     } 
} 
%
%
%
\maketitle
\section{Introduction}
\label{intro}

  Elliptic flow is expected to be one of the key observables to study an early stage 
  of heavy ion collisions \cite{Ollitrault:1992bk}. 
  It is defined by the second harmonic Fourier coefficient
  \begin{equation}
    v_2 = \left<\cos{(2[\phi - \Psi_{\mathrm{RP}}])}\right>,
  \end{equation}
  where $\phi$ is the azimuthal angle of emitted particles, $\Psi_{\mathrm{RP}}$
  is the azimuthal angle of reaction plane and brackets denote the average over all 
  particles and events. 

  The PHENIX experiment at Relativistic Heavy Ion Collider (RHIC)
  have measured the $v_2$ for 
  identified charged hadrons \cite{Adler:2003kt,Adare:2006ti},
  $\phi$ mesons and deuterons \cite{Afanasiev:2007tv},
  $\pi^0$'s and photons \cite{Adler:2005rg}
  as well as electrons from heavy flavor decays \cite{Adler:2005ab}
  at midrapidity.
  The mass ordering of $v_2$ for identified hadrons were qualitatively 
  explained by ideal hydrodynamics in transverse momentum 
  $p_{\mathrm{T}} <$ 2 GeV/$c$ \cite{Adler:2003kt}. 
  For intermediate $p_{\mathrm{T}} = 2 - 6$ GeV/$c$,
  a universal parton $v_2$ were obtained by dividing $v_2$ and $p_{\mathrm{T}}$ by 
  constituent quarks for each hadron \cite{Adler:2003kt,Adare:2006ti}.
  The $v_2$ for $\phi$ meson was also found to follow the quark number scaling, 
  which support that the parton $v_2$ have already developed prior to the 
  hadronization \cite{Afanasiev:2007tv}.
  Because the cross section of $\phi$ meson to non-strange hadrons are small, 
  $\phi$ meson $v_2$ is less sensitive to the late hadronic stage.
  The finite $v_2$ for electrons from heavy flavor decays implied 
  the non-zero charm $v_2$ \cite{Adler:2005ab}. Comparison of $v_2$
  with transport model calculation suggest that the viscosity to 
  entropy density ratio is close to the quantum lower bound 1/4$\pi$ \cite{Adare:2006nq}.

  These measurements were done by using an event plane 
  determined from the Beam-Beam Counter (BBC) located at pseudorapidity 
  $|\eta|$ = 3.1 - 3.9. The large pseudorapidity separation $|\Delta\eta| \sim 3$
  from midrapidity would reduce non-flow effects, 
  which are correlations not originated from the reaction plane 
  such as jets, resonance decays and so on.
  Fluctuations of $v_2$ were also considered as the non-flow contributions \cite{Adler:2002pu}, 
  which would become more important in smaller system, such as Cu + Cu collisions.
  It is crucial to understand how non-flow contributions 
  affect the event plane determined at the BBC so that
  we could validate a sensitivity to real collective flow 
  on our measured $v_2$ based on the BBC.

  In this paper, we present the $v_2$ with respect to the event plane 
  from directed flow determined at the Shower Maximum Detector (SMD),
  which is located at $|\eta| > 6$. The larger rapidity separation 
  could reduce the possible non-flow effects on our measured $v_2$.
  We will compare the $v_2$ results from the event planes determined 
  at the BBC and SMD and discuss the possible non-flow contributions 
  on the $v_2$. 

\section{Data Analysis}
\label{sec:analysismethod}

  In this study, we analyzed $\sim$ 650 M events collected by the PHENIX 
  experiment in Au + Au at $\sqrt{s_{NN}}$ = 200 GeV. 
  Minimum bias events were selected within a collisions z-vertex $\pm$ 30 cm.
  Event centrality was determined by the correlation between the energy 
  deposit at the Zero Degree Calorimeter (ZDC) and number of charged particles 
  at the BBC. 
  Tracking were done by the Drift Chamber (DC) and Pad Chambers (PCs) at 
  the central arm $|\eta| < 0.35$. Transverse momentum were determined 
  by the incident angle at the DC. The polar angle of the tracks were 
  obtained by the hit at the inner PC (PC1) and the collision vertex 
  from the BBC. Track associations were made by comparing hit positions 
  with the projection of the DC tracks to the outer Pad Chamber (PC3).
  Tracks were required to have a hit on the PC3 within $\pm$ 3 $\sigma$
  of the expected hit location in both azimuthal and beam directions.
  Large energy deposit $E/p > 0.2$ at the Electromagnetic Calorimeter (EMCal) 
  were also required, where $E$ is the energy deposit in EMCal and $p$ is 
  the momentum determined at the DC. 
  It was necessary to eliminate the background mainly from photon conversions 
  and resonance decays so as to improve the signal to background ratio for $p_{\mathrm{T}} > 4$ GeV/$c$.

  The $v_2$ was measured by an event plane method \cite{Poskanzer:1998yz}
  and was obtained by dividing the measured $v_2$ by the event plane 
  resolution
  \begin{equation}
    v_2 = \frac{v_2^{obs}}{\mathrm{Res}\{\Psi_{n}\}} = \frac{\left<\cos{(2[\phi-\Psi_{n}])}\right>}
    {\left<\cos{(2[\Psi_n - \Psi_{\mathrm{RP}}])}\right>},
  \end{equation}
  where $\phi$ is the azimuth of charged hadrons at the central arm ($|\eta| < 0.35$),
  $\Psi_n$ is the event plane from the $n$-th harmonic flow
  ($n$ = 1 for the ZDC-SMD, $n$ = 2 for the BBC)
  and $v_2^{obs}$ is the measured $v_2$ with respect to the event plane $\Psi_n$.
  Event planes were determined from the $v_2$ at the BBC and the central arm 
  as well as the directed flow $v_1 = \left<\cos{(\phi - \Psi_{\mathrm{RP}})}\right>$ 
  at the Shower Maximum Detector (SMD).
  The central arm event plane is only used to evaluate the event plane 
  resolutions.
  The SMDs are located at the same acceptance of the ZDCs, $|\eta| > 6$, 
  and measure transverse positions of spectator neutrons.
  The measured $v_2$'s are denoted as $v_2$\{ZDC-SMD\} and $v_2$\{BBC\} 
  for the ZDC-SMD and BBC event planes, respectively.

  The event plane determined at the ZDC-SMD can minimize 
  non-flow correlations as well as $v_2$ fluctuations 
  because of the following reasons. 
  First, the pseudorapidity gap from midrapidty is 6, 
  which is higher than what we have previously studied the $v_2$ 
  by using the BBC. Second, the ZDC-SMD event plane is 
  determined from directed flow. This mixed harmonic method 
  involves three particle correlations and thus direct 
  two particle correlations, which is dominant contributions 
  from non-flow effects, do not affect the measured $v_2$.
  Third, ZDC-SMD measures spectator neutrons rather than 
  participants. Therefore, $v_2$ fluctuations are suppressed 
  up to the fluctuation of spectator neutrons.

  \begin{figure}[tb]
  \resizebox{1.0\linewidth}{!}{%
    \includegraphics{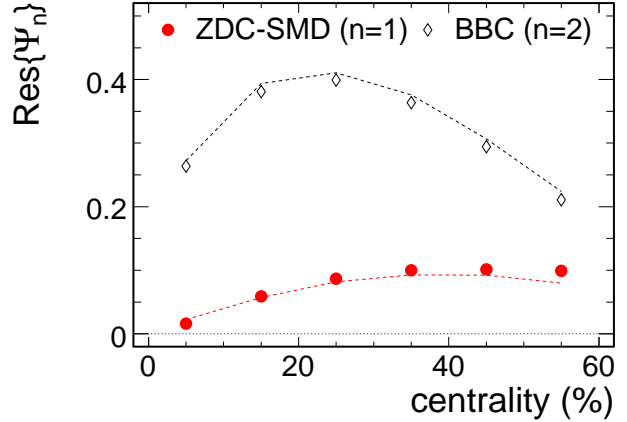}
  }
  \caption{
    Event plane resolutions as a function of centrality
    for the ZDC-SMD event plane (solid circles) and 
    the BBC (open diamonds) by Eq.~(\ref{eq:resolution_2}). 
    Dashed lines represent the resolutions calculated by Eq.~(\ref{eq:resolution_3}).
  }
  \label{fig:fig1}
  \end{figure}
    Fig.~\ref{fig:fig1} shows the event plane resolutions 
  as a function of centrality. At least two independent 
  event planes are required in order to evaluate the resolution
  since the azimuth of true reaction plane is unknown.
  The resolution from two independent event planes is calculated by 
  \begin{equation}
    \mathrm{Res}\{\Psi_n\} = C\sqrt{\left<\cos{(2[\Psi_n^{-}-\Psi_{n}^+])}\right>},
    \label{eq:resolution_2}
  \end{equation}
  where $\Psi_n^{+}$ and $\Psi_n^{-}$ denote the event planes determined 
  at the forward and backward pseudorapidities, respectively. 
  A constant parameter $C$ is very close to $\sqrt{2}$ for both BBC and ZDC-SMD 
  due to low resolution \cite{Poskanzer:1998yz}.  
  The ZDC-SMD resolution is about factor 4 smaller than that of BBC 
  because the ZDC-SMD event plane is determined from directed flow.
  It is approximately proportional to $v_1^2M^{\mathrm{SMD}}$, 
  where $M^{\mathrm{SMD}}$ is multiplicity used to determine the 
  ZDC-SMD event plane, whereas the BBC resolution 
  is roughly proportional to $v_2\sqrt{M^{\mathrm{BBC}}}$.

  The resolutions were also evaluated by adding a reference event plane
  \begin{equation}
    \mathrm{Res}\{\Psi_n\} = \sqrt{ 
      \frac{\left<\cos{(2[\Psi_l^A-\Psi_{n}])}\right>\left<\cos{(2[\Psi_n-\Psi_m^B])}\right>}
    {\left<\cos{(2[\Psi_m^B - \Psi_l^A])}\right>}
    },
    \label{eq:resolution_3}
  \end{equation}
  where $l$, $m$ and $n$ denote the harmonics for event plane $\Psi^A$, $\Psi^B$ 
  and $\Psi$, respectively. Dashed lines in Fig.~\ref{fig:fig1} show the 
  resolutions calculated by Eq.~(\ref{eq:resolution_3}). For example, 
  the BBC resolution was calculated by inserting
  $\Psi_n$ = $\Psi_2^{\mathrm{BBC}}$,
  $\Psi_l^A$ = $\Psi_1^{\mathrm{ZDC-SMD}}$,
  and $\Psi_m^B$ = $\Psi_2^{\mathrm{CNT}}$
  where CNT denote the central arm.
  One can find that the dashed lines are systematically 
  lower for the ZDC-SMD, and higher for the BBC.
  The comparison of $v_2$ from two different 
  resolutions will be presented in the next section.

\section{Results}
\label{sec:results}

  We will present the preliminary results of $v_2$\{BBC\} as well as \{ZDC-SMD\}
  in Au + Au collisions at $\sqrt{s_{NN}}$ = 200 GeV measured at the PHENIX experiment.
  Section \ref{subsec:sec3.1} will give comparison of the $v_2$ between BBC and 
  ZDC-SMD event planes. Results between PHENIX and STAR experiments will be 
  compared in Section \ref{subsec:sec3.2}.
  In Section \ref{subsec:sec3.3}, centrality dependence of the $v_2$\{ZDC-SMD\} 
  will be compared with the $v_2$\{BBC\}.

  \subsection{\label{subsec:sec3.1} Comparison of $v_2$\{BBC\} with $v_2$\{ZDC-SMD\}}

  \begin{figure}[tb]
  \resizebox{1.0\linewidth}{!}{%
    \includegraphics{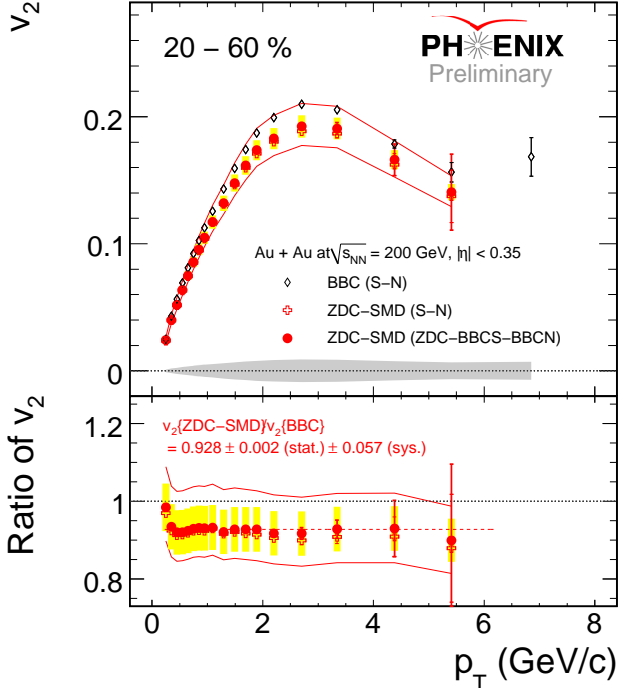}
  }
  \caption{
    (Top) Comparison of $v_2$ as a function of $p_{\mathrm{T}}$ in 20 - 60 \% centrality
    for the BBC (open diamonds), the ZDC-SMD from two different event plane resolutions 
    (open crosses and solid circles, see texts).
    Gray bands, solid red lines and yellow boxes represent systematic 
    uncertainties on the $v_2$\{BBC\} and $v_2$\{ZDC-SMD\}.
    (Bottom) The ratio of $v_2$\{ZDC-SMD\} to $v_2$\{BBC\}
    as a function of $p_{\mathrm{T}}$. Dashed line denote the fit result by constant.
  }
  \label{fig:fig2}
  \end{figure}

    Fig.~\ref{fig:fig2} shows the $v_2$\{ZDC-SMD\} as a function of $p_{\mathrm{T}}$
  in 20 - 60 \% centrality. For comparison, the $v_2$\{BBC\} is also plotted by 
  open diamonds. The $v_2$ increases linearly up to $p_{\mathrm{T}} \sim$ 3 GeV/$c$,
  reach maximum $\sim$ 0.2 and then start decreasing for higher $p_{\mathrm{T}}$.
  The $v_2$\{ZDC-SMD\} (S-N), which is obtained from the resolution in Eq.~(\ref{eq:resolution_2}),
  is about 7 \% systematically lower than the $v_2$\{BBC\}, 
  while the results are consistent within systematic uncertainties. 
  We also plot the $v_2$\{ZDC-SMD\} (ZDC-BBCS-BBCN) as shown by solid circles,
  which is obtained from the resolution in Eq.~(\ref{eq:resolution_3}) by 
  inserting 
  $\Psi_n = \Psi_1^{\mathrm{ZDC-SMD}}$,
  $\Psi_n^A = \Psi_2^{\mathrm{BBCS}}$ and
  $\Psi_n^B = \Psi_2^{\mathrm{BBCN}}$.
  The BBCS and BBCN denote the backward and forward BBC, respectively.
  The $v_2$\{ZDC-SMD\} from two different resolutions are in good agreement 
  within systematic uncertainties.
  Bottom panel shows the ratio of $v_2$\{ZDC-SMD\} to $v_2$\{BBC\} 
  as a function of $p_{\mathrm{T}}$. One can see that 
  the ratio is constant within systematic errors
  in the measured $p_{\mathrm{T}}$ range.

  \subsection{\label{subsec:sec3.2} PHENIX vs STAR}

  \begin{figure}[tb]
  \resizebox{1.0\linewidth}{!}{%
    \includegraphics{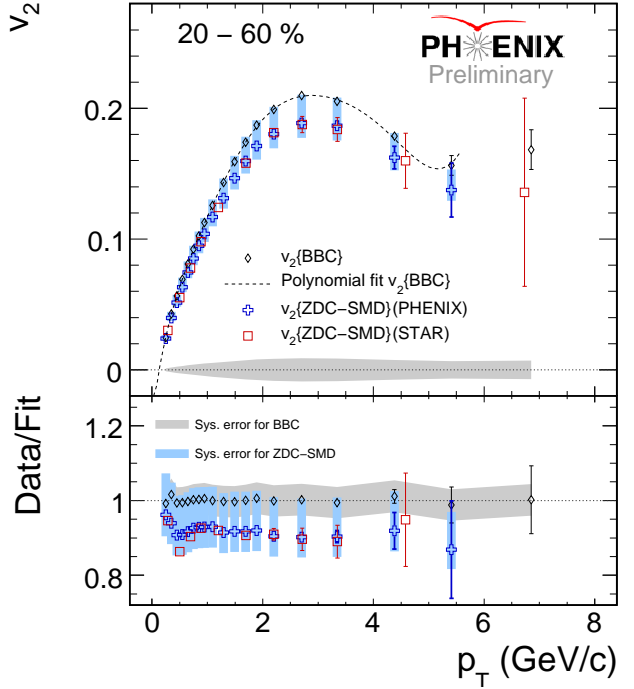}
  }
  \caption{
    Comparison of the $v_2$\{ZDC-SMD\} between PHENIX and STAR experiments
    in 20 - 60 \% centrality. The STAR $v_2$\{ZDC-SMD\} is taken from \cite{Wang:2006}.
    Gray bands and light blue boxes denote the systematic uncertainties 
    on the $v_2$\{BBC\} and $v_2$\{ZDC-SMD\}, respectively.
    Bottom panel shows the ratio of $v_2$ to the fourth polynomial fit of the 
    $v_2$\{BBC\} as a function of $p_{\mathrm{T}}$.
  }
  \label{fig:fig3}
  \end{figure}
  
  Fig.~\ref{fig:fig3} show the comparison of the PHENIX $v_2$\{ZDC-SMD\} 
  with STAR result \cite{Wang:2006} in 20 - 60 \% centrality bin. 
  Only statistical errors are shown for the STAR $v_2$.  
  Both PHENIX and STAR results are obtained by the resolution in Eq.~(\ref{eq:resolution_2})
  and thus the results of $v_2$\{ZDC-SMD\} are extracted by the exactly 
  same method.
  Data symbols (open diamonds and open crosses) are the same as shown 
  in Fig.~\ref{fig:fig2}.
  For a quantitative comparison, the ratio of $v_2$ to the $v_2$\{BBC\}
  is plotted in bottom panel in Fig.~\ref{fig:fig3}.
  The denominator of the ratio is fitting result of the $v_2$\{BBC\} 
  by fourth polynomial function.
  One can see that the results agree very well within systematic 
  errors.

  \subsection{\label{subsec:sec3.3} Centrality dependence of $v_2$}

  \begin{figure*}[htbp]
    \subfigure[Without CNT event plane resolution]{
      \resizebox{0.50\textwidth}{!}{ \includegraphics{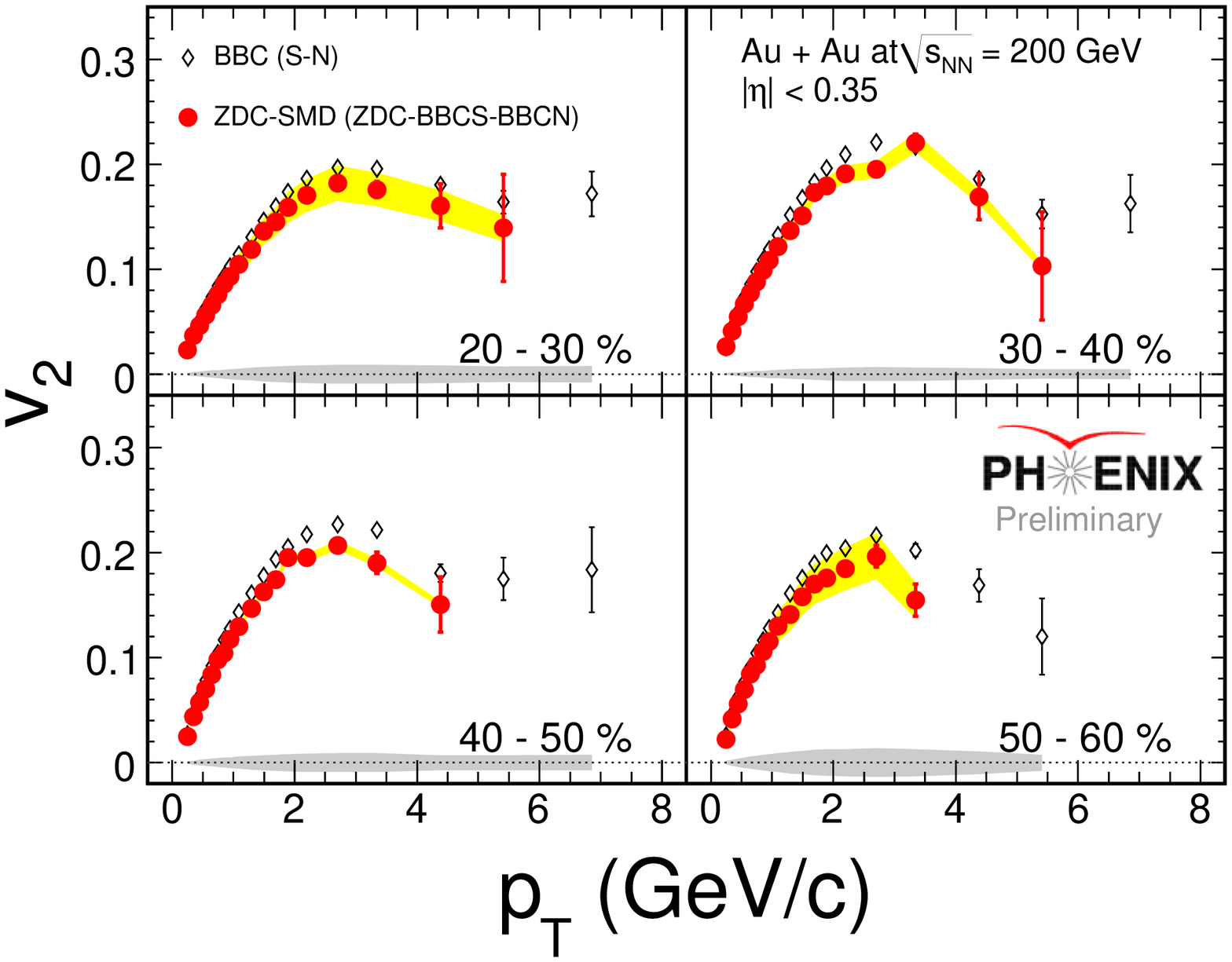} }
    }
    \hspace{0.005\textwidth}
    \subfigure[With CNT event plane resolution]{
      \resizebox{0.50\textwidth}{!}{ \includegraphics{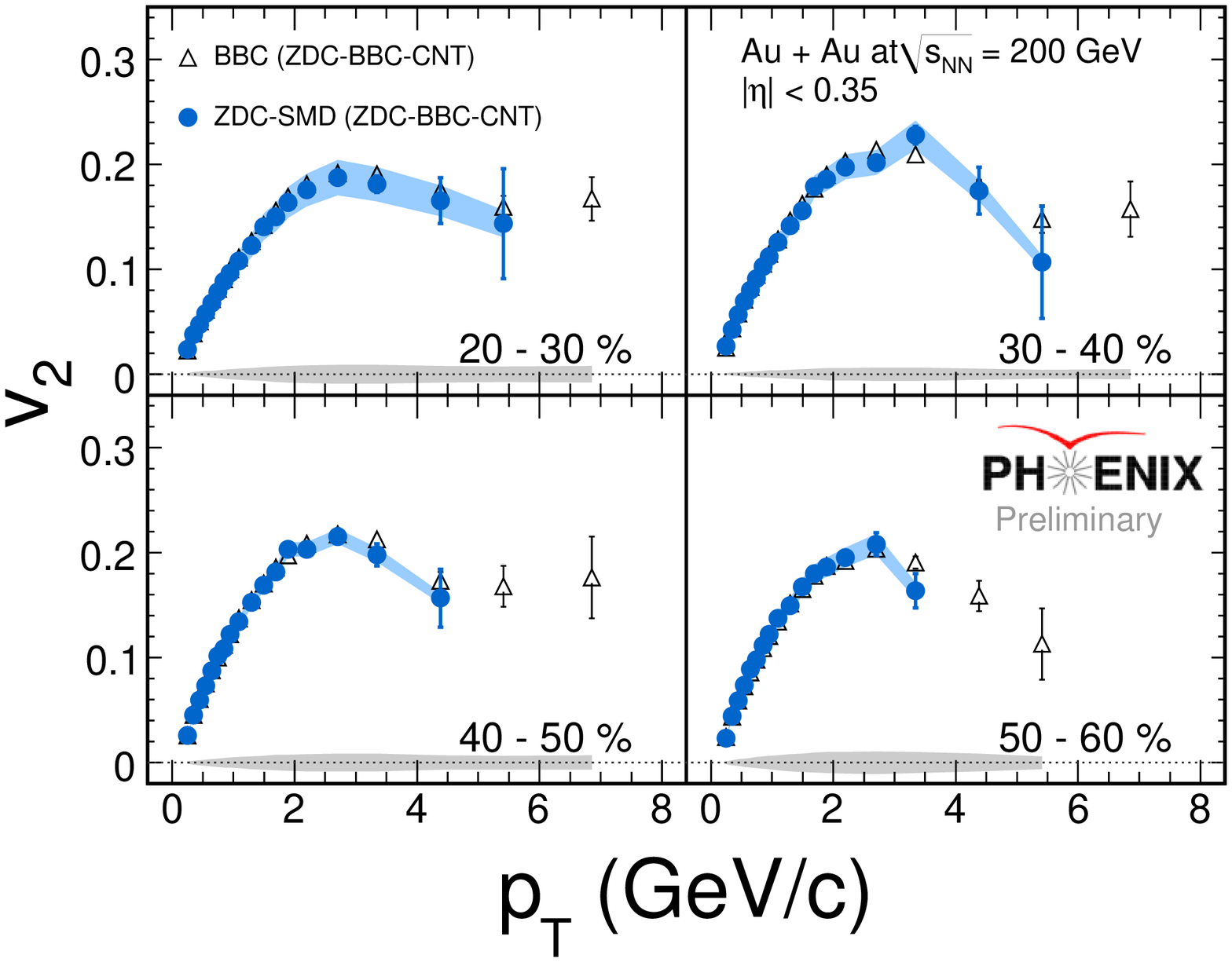} }
    }
    \caption{ \label{fig:fig4-5}
      Comparison of $v_2$\{ZDC-SMD\} to $v_2$\{BBC\} as a function of $p_{\mathrm{T}}$ for 
      10 \% step centrality bin in 20 - 60 \%.
      Left (a) and right (b) figures show the results without and with the central arm (CNT) 
      event plane resolution, respectively.
    }
  \end{figure*}

    Fig.~\ref{fig:fig4-5} show the $v_2$($p_{\mathrm{T}}$) for 10 \% step 
  centrality bin in 20 - 60 \%.  In left figures, the $v_2$\{ZDC-SMD\} 
  is essentially consistent with the $v_2$\{BBC\} within systematic errors.
  In peripheral 40 - 60 \%, we find that the $v_2$\{ZDC-SMD\} is 5 - 10 \% 
  lower than the $v_2$\{BBC\}. The lower $v_2$\{ZDC-SMD\} could suggest 
  possible non-flow contributions on the $v_2$\{BBC\}.

  In right figures, both BBC and ZDC-SMD resolutions are calculated 
  by ZDC-SMD, BBC and CNT combinations in Eq.~(\ref{eq:resolution_3}).
  Since the non-flow effects are expected to be maximum at the midrapidity,
  the CNT event plane could have maximal sensitivity to the 
  non-flow contributions. 
  Therefore, by including the CNT event plane resolution, 
  we could study how non-flow contributions on the CNT resolution 
  modify the $v_2$\{BBC\} as well as $v_2$\{ZDC-SMD\}.
  We find that the $v_2$\{ZDC-SMD\} become more consistent with the 
  $v_2$\{BBC\} by including the CNT event plane resolution.
  Because the ZDC-SMD (BBC) resolution from the ZDC-BBC-CNT combination 
  in Eq.~(\ref{eq:resolution_3}) is lower (higher) than that from Eq.~(\ref{eq:resolution_2})
  as shown in Fig.~\ref{fig:fig1}. 
  Resulting $v_2$\{ZDC-SMD\} ($v_2$\{BBC\}) become higher (lower) than 
  those from the resolution by forward and backward correlations.
  Therefore, the $v_2$\{ZDC-SMD\} is closer to the $v_2$\{BBC\} when 
  one include CNT event plane to evaluate the resolution.

\section{Conclusion}
\label{sec:conclusion}

  In summary, we have measured unidentified charged hadron elliptic 
  flow with respect to the ZDC-SMD event plane from directed flow
  in Au + Au at $\sqrt{s_{NN}}$ = 200 GeV. 
  The $v_2$\{ZDC-SMD\} was compared with the $v_2$ measured with respect 
  to the event plane determined at the BBC.
  We found that the $v_2$\{ZDC-SMD\} was basically consistent with 
  the $v_2$\{BBC\} within systematic uncertainties in 20 - 60 \% centrality.
  Several different choice of event plane resolutions were studied. 
  We found that resulting $v_2$\{ZDC-SMD\} was still consistent with 
  the $v_2$\{BBC\} even if the CNT event plane was included in the 
  event plane resolution. 
  The difference of $v_2$ between BBC and ZDC-SMD event plane, $\sim$ 5 - 10 \%, 
  at 40 - 60 \% centrality bins could attribute to the possible 
  non-flow effects.
  This result indicate that the non-flow effects are essentially 
  minimal on the $v_2$\{BBC\} in 20 - 60 \% centrality
  because the $v_2$\{ZDC-SMD\} is expected to be unbiased by the non-flow 
  contributions.

%
%

%
%


\end{document}